\documentclass[fleqn,twoside]{article}
\usepackage{espcrc2}

\usepackage{graphicx}
\usepackage[figuresright]{rotating}

\newcommand{\AmS}{{\protect\the\textfont2
  A\kern-.1667em\lower.5ex\hbox{M}\kern-.125emS}}

\def\be{\begin{equation}}
\def\ee{\end{equation}}
\def\bea{\begin{eqnarray}}
\def\eea{\end{eqnarray}}
\def\beq{\begin{equation}}
\def\eeq{\end{equation}}
\def\bq{\begin{quote}}
\def\eq{\end{quote}}
\def\gappeq{\mathrel{\rlap {\raise.5ex\hbox{$>$}} {\lower.5ex\hbox{$\sim$}}}}
\def\lappeq{\mathrel{\rlap{\raise.5ex\hbox{$<$}} {\lower.5ex\hbox{$\sim$}}}}

\def\epm#1#2{\hbox{${\lower1pt\hbox{$\scriptstyle +#1$}}
\atop {\raise1pt\hbox{$\scriptstyle -#2$}}$}}

\def\gsim{\mathrel{\rlap{\lower4pt\hbox{\hskip1pt$\sim$}}
    \raise1pt\hbox{$>$}}}         

\def\frac#1#2{{{#1}\over {#2}}}

\def\bq{\bar{q}}
\catcode`@=11 
\def\slash#1{\mathord{\mathpalette\c@ncel#1}}
 \def\c@ncel#1#2{\ooalign{$\hfil#1\mkern1mu/\hfil$\crcr$#1#2$}}
\def\lsim{\mathrel{\mathpalette\@versim<}}
\def\gsim{\mathrel{\mathpalette\@versim>}}
 \def\@versim#1#2{\lower0.2ex\vbox{\baselineskip\z@skip\lineskip\z@skip
       \lineskiplimit\z@\ialign{$\m@th#1\hfil##$\crcr#2\crcr\sim\crcr}}}
\catcode`@=12 

\title{Neutrino 2004: Concluding Talk}

\author{Guido Altarelli
{CERN, Department of Physics, Theory Division\\
CH--1211 Geneva 23, Switzerland}%
		}
       
\begin{document}

\begin{abstract}
We review the  highlights of Neutrino 2004 and recall the main lessons from neutrinos in recent years 
that had a great impact on particle physics and cosmology.\\
\vspace{1pc}
\\
CERN-PH-TH/2004-197
\end{abstract}

\maketitle
\section{Introduction}

I have chosen the formula "Concluding Talk" in the title of my contribution to Neutrino 2004. Indeed this is not a "Summary Talk": it is impossible to review in 30 minutes/10 pages the great variety of results and ideas and projects for the future that have been presented at this one-week-long Conference. Also, I am not really competent on some areas, like, for example, detector technology, which was the subject of many impressive talks. It is not a "Conclusion" talk either: the subject of neutrinos is at present a gigantic work in progress and we are not at the end of a particular phase where a sharp line can easily be drawn. Rather, in my presentation I will list the new experimental results that most impressed me at this Conference, I will put in the general context of fundamental physics the great additions to our knowledge that in recent years were obtained from neutrinos, at present one of the most dynamic domains, and discuss their implications for particle physics and cosmology. I will mainly restrict references to only mentioning the talks given at this Conference where the different ideas, results and projects were presented and discussed. 

\section{New Data}

By now there is convincing evidence for solar and atmospheric neutrino oscillations (while the LSND signal remains unconfirmed and is being investigated by MiniBoone [S. Brice]). The $\Delta m^2$ values and mixing angles are known with fair accuracy [S. Goswami]. A summary, taken from Ref. \cite{one}, of the results known before this conference is shown in Table~\ref{tab01}. For the $\Delta m^2$ we have:  $\Delta m^2_{atm}\sim 2.5~10^{-3}~eV^2$ and  $\Delta m^2_{sol}\sim 8~10^{-5}~eV^2$. As for the mixing angles, two are large and one is small. The atmospheric angle $\theta_{23}$ is large, actually compatible with maximal but not necessarily so: at $3\sigma$: $0.31 \lappeq \sin^2{\theta_{23}}\lappeq 0.72$ with central value around  $0.5$. The solar angle $\theta_{12}$ is large, $\sin^2{\theta_{12}}\sim 0.3$, but certainly not maximal (by more than 5$\sigma$). The third angle $\theta_{13}$, strongly limited mainly by the CHOOZ experiment, has at present a $3\sigma$ upper limit given by about $\sin^2{\theta_{13}}\lappeq 0.08$ (it moved up a bit recently).

\begin{table*}[htb]
\begin{center}
\caption[]{Square mass differences and mixing angles\cite{one}}
\label{tab01}
\begin{tabular}{|c|c|c|c|}   
\hline  
& & & \\   
&{\tt lower limit} & {\tt best value} & {\tt upper limit}\\
&($3\sigma$)& & ($3\sigma$)\\
\hline
& & & \\
$(\Delta m^2_{sun})_{\rm LA}~(10^{-5}~{\rm eV}^2)$ & 5.4& 6.9&9.5\\
& & & \\
\hline
& & & \\
$\Delta m^2_{atm}~(10^{-3}~{\rm eV}^2)$ & 1.4& 2.6& 3.7\\
& & & \\
\hline
& & & \\
$\sin^2\theta_{12}$ & 0.23 & 0.30 &0.39\\
& & & \\
\hline
& & & \\
$\sin^2\theta_{23}$ & 0.31 & 0.52 &0.72\\
& & & \\
\hline
& & & \\
$\sin^2\theta_{13}$ & 0 & 0.006 &0.054\\
& & & \\
\hline
\end{tabular} 
\end{center}
\end{table*}

A most interesting development at this Conference has been the release of the new results from the KamLAND experiment [G. Gratta], which has confirmed, using $\bar \nu_e$'s from many nuclear reactors,  the oscillation interpretation of the solar neutrino deficit. KamLAND has a tremendous sensitivity to $\Delta m^2_{sol} \equiv \Delta m^2_{12}$ while it does not constrain $\sin^2{\theta_{12}}$ very much. The combined solar $\nu$ plus KamLAND 2-flavour analysis results are  $\Delta m^2_{12}\sim (8.2+0.6-0.5) ~10^{-5}~eV^2$ and $\sin^2{\theta_{12}}\sim 0.29 \pm 0.04$ (see Fig. 1) .

\begin{figure*}[htb]
\includegraphics[width=16cm]{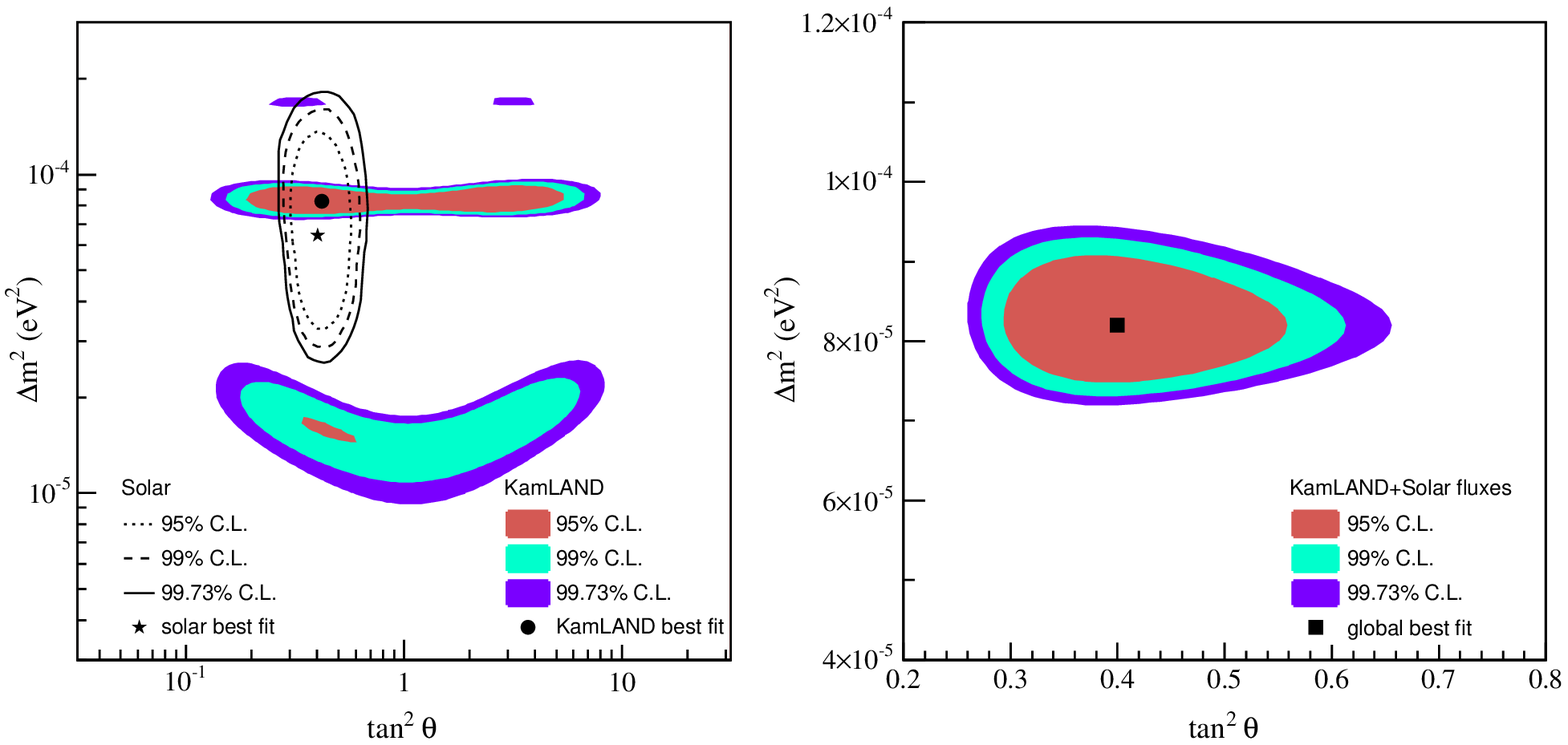}
\caption{Allowed regions of neutrino oscillation parameters from KamLAND [G.Gratta]}
\end{figure*}

 It is very remarkable that in the last few years, thanks to the SNO [J. Wilkerson] and KamLAND experiments, not only the solar neutrino interpretation in terms of oscillations has been confirmed, but the MSW-LA solution has been selected and the value of $\Delta m^2_{12}$ is now precisely determined, while it was previously ambiguous by several orders of magnitude (if one moved from one solution to another). It is also interesting that a first hint of the direct observation of the oscillation pattern is visible in the KamLAND "L"/E distribution (see Fig.~2), where the "L" is in quote because the distance of the different reactors is different and only an average distance can be defined. 
 
\begin{figure}[htb]
\begin{center}
\includegraphics[angle=270,width=8cm]{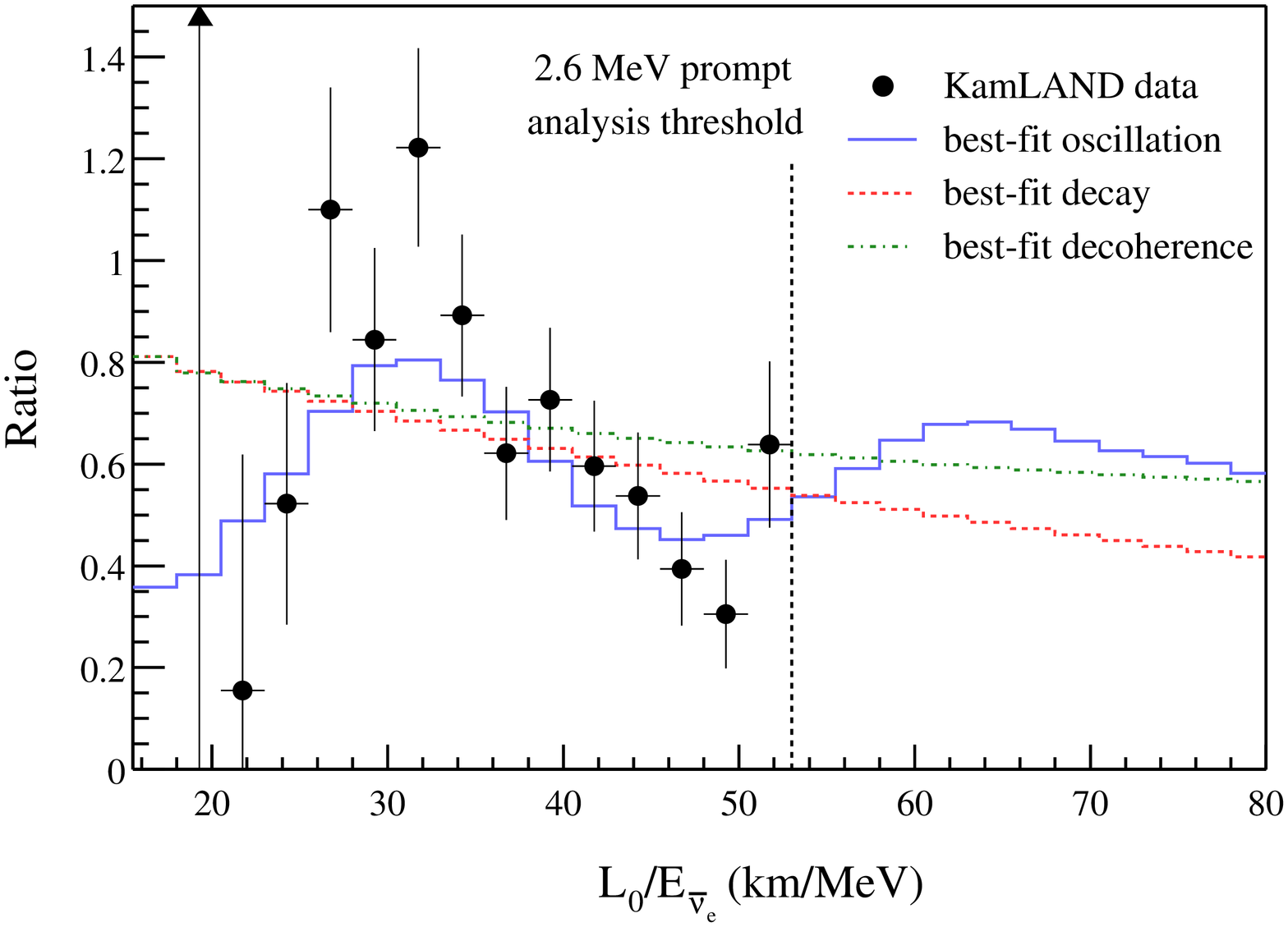}
\caption{The $L_0/E$ KamLAND spectrum data compared with the best fit oscillation curve and with the decay and decoherence models [G. Gratta]}
\end{center}
\end{figure}

\begin{figure}[htb]
\begin{center}
\includegraphics[width=8cm]{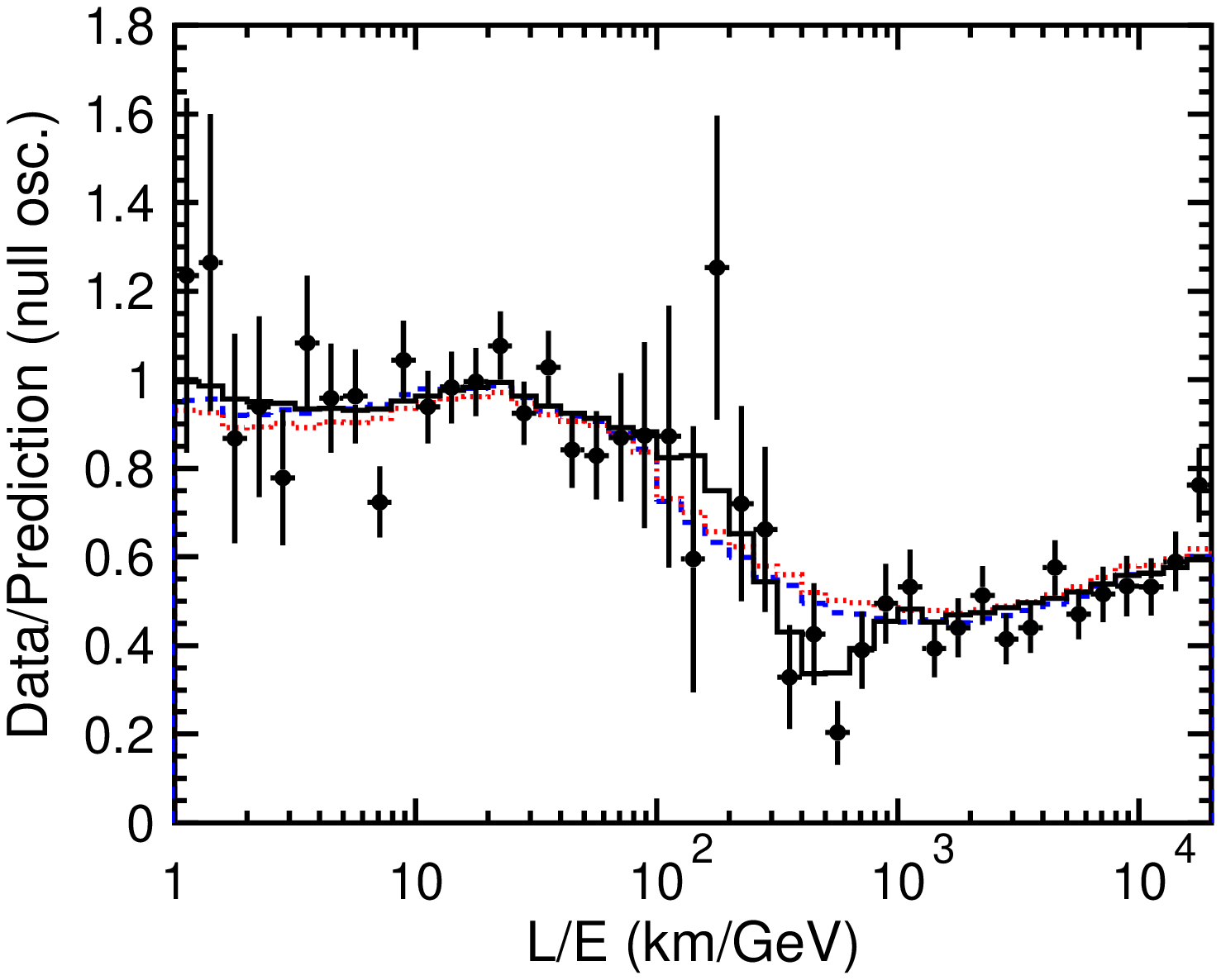}
\caption{Ratio of the SuperKamiokande data to the MC events without neutrino oscillations (points) vs. $L/E$ compared with the $\nu$-oscillation expectation (solid) and those from the decay and decoherence models [E. Kearns]}
\end{center}
\end{figure}

For atmospheric neutrinos the main developments discussed at Neutrino 2004 were the results of the SuperKamiokande  (SK)
L/E analysis [E. Kearns] and the reported progress of the K2K experiment [T. Nakaya]. The SK L/E analysis leads to the indication of an oscillation dip seen at $\sim 500 Km/GeV$ in the L/E distribution (see Fig.~3), which is also a direct hint of the oscillation pattern but in this case for atmospheric neutrinos. The L/E analysis, which is a different way of looking at the SK data, leads to consistent values of $\Delta m^2_{23}$ and $\sin^2{\theta_{23}}$ as from the ordinary analysis. However, a stronger lower bound on $\Delta m^2_{23}$ is obtained from the L/E analysis. The K2K experiment is an accelerator experiment aimed at a terrestrial confirmation of the atmospheric neutrino oscillations. It suffered an interruption due to the unfortunate incident that ended SK I in 2001. Now resumed, with $8.9~10^{19}$ POT, K2K has confirmed atmospheric $\nu$ oscillations at $3.9\sigma$: arising from combining the rate and the spectrum observations. From the rate of $\nu_{\mu}$ disappearance ($N_{EXP} =150.9+11.6-10.0$ is the number of events expected from no oscillations and $N_{OBS} =108$ is the number of observed events) oscillations are indicated at $2.9\sigma$ and from the distortion of the $E_{\nu}$ spectrum at $2.5\sigma$. The values of the fitted parameters from K2K are in perfect agreement with those obtained from atmospheric oscillations (central values: $\Delta m^2_{23} = 2.73~10^{-3}~eV^2$ and maximal $\theta_{23})$.

\section{Neutrino Masses}

Neutrino oscillations measure $\Delta m^2$ but do not provide information about the absolute neutrino spectrum and cannot distinguish between
pure Dirac and Majorana neutrinos. The following experimental information on the absolute scale of neutrino masses is available. From the endpoint of tritium beta decay spectrum we have  an absolute upper limit of
2.2 eV (at 95\% C.L.) on the mass of "$\bar \nu_e$" [K. Eitel], which, combined with the observed oscillation
frequencies under the assumption of three CPT-invariant light neutrinos, represents also an upper bound on the masses of
all active neutrinos. Complementary information on the sum of neutrino masses is also provided by the galaxy power
spectrum combined with measurements of the cosmic  microwave background anisotropies. According to the recent analysis of
the WMAP collaboration  [O. Lahav], 
$\sum_i \vert m_i\vert < 0.69$ eV (at 95\% C.L.).  More conservative analyses give $\sum_i \vert
m_i\vert < 1- 1.8$ eV, still much more restrictive than the laboratory bound. But, to some extent, the cosmological bound depends on a number of assumptions (or, in fashionable terms, priors). However, for 3 degenerate neutrinos of mass $m$, depending on our degree of confidence in the cosmological bound, we can be sure that $m \lappeq
0.23-0.3-0.6~eV$.
The discovery of $0\nu \beta \beta$ decay would be very important because it would directly establish lepton number violation and
the Majorana nature of $\nu$'s, and provide direct information on the absolute
scale of neutrino masses [J. Vergados, J. Suhonen, F. Avignone].  
The present limit from $0\nu \beta \beta$ is [H.V. Klapdor-Kleingrothaus] $\vert m_{ee}\vert< 0.2$ eV or to be
more conservative
$\vert m_{ee}\vert < (0.3\div 0.5)$ eV, where $ m_{ee} =  \sum{U_{ei}^2 m_i}$ in terms of the mixing matrix and the mass eigenvalues (see eq.(\ref{3nu1gen}). Possible evidence of a signal near the upper bound was discussed by Klapdor-Kleingrothaus.

Given that neutrino masses are certainly extremely
small, it is really difficult from the theory point of view to avoid the conclusion that L conservation must be violated.
In fact, in terms of lepton number violation the smallness of neutrino masses can be naturally explained as inversely proportional
to the very large scale where L is violated, of order $M_{GUT}$ or even $M_{Pl}$.
In the see-saw mechanism [P. Binetruy] the resulting neutrino mass matrix reads:
\beq  m_{\nu}=m_D^T M^{-1}m_D~~~.
\eeq that is the light neutrino masses are quadratic in the Dirac
masses (or a different parameter of the same order) and inversely proportional to the large Majorana mass.  For
$m_{\nu}\approx \sqrt{\Delta m^2_{atm}}\approx 0.05$ eV and 
$m_{\nu}\approx m_D^2/M$ with $m_D\approx v
\approx 200~GeV$ we find $M\approx 10^{15}~GeV$ which indeed is an impressive indication for
$M_{GUT}$. Thus neutrino masses are a probe into the physics at $M_{GUT}$. By now GUT's are part of our culture in particle physics and most of us believe that Grand Unification must be a feature of the final theory. But the very large value of $M_{GUT}$ close to $M_{Pl}$ poses the problem of the relation of $m_W$ (or the weak scale) with $M_{GUT}$-$M_{Pl}$ known as the hierarchy problem: the SM is unstable versus quantum fluctuations and new physics near the weak scale should be present to stabilize the theory or an enormous fine tuning would be required.

\section{Neutrino Masses and New Physics Scenarios}

So the hierarchy problem demands new physics to be very close (in
particular the mechanism that quenches the top loop contribution to the Higgs mass). Actually, this new physics must be rather special, because it must be
very close, yet its effects are not clearly visible (the "LEP Paradox"). Examples of proposed classes of solutions
for the hierarchy problem are:

¥ Supersymmetry. In the limit of exact boson-fermion symmetry the quadratic divergences of bosons cancel so that
only log divergences remain. However, exact SUSY is clearly unrealistic. For approximate SUSY,
which is the basis for all practical models, $\Lambda$ is replaced by the splitting of SUSY multiplets, $\Lambda\sim
m_{SUSY}-m_{ord}$. In particular, the top loop is quenched by partial cancellation with s-top exchange. 

¥ Technicolor. The Higgs system is a condensate of new fermions. There is no fundamental scalar Higgs sector, hence no
quadratic devergences associated to the $\mu^2$ mass in the scalar potential. This mechanism needs a very strong binding force,
$\Lambda_{TC}\sim 10^3~\Lambda_{QCD}$. It is  difficult to arrange that such nearby strong force is not showing up in
precision tests. Hence this class of models has been disfavoured by LEP, although some special class of models have been devised aposteriori, like walking TC, top-color assisted TC etc. 

¥ Large~compactified~extra~dimensions. The idea is that $M_{Pl}$ appears very large, or equivalently that gravity seems very weak,
because we are fooled by hidden extra dimensions so that the real gravity scale is reduced down to
$o(1~TeV)$. This possibility is very exciting in itself and it is really remarkable that it is compatible with experiment. Many different subclasses of models belong to this general framework.

¥ "Little~Higgs" models. In these models extra symmetries allow $m_h\not= 0$ only at two-loop level, so that $\Lambda$
can be as large as
$o(10~TeV)$ with the Higgs mass within present bounds (the top loop is quenched by exchange of heavy vectorlike new charge-2/3 quarks).

What is really unique to SUSY with respect to all other extensions of the SM listed above is that the MSSM or
similar models are well defined and computable up to $M_{Pl}$ and, moreover, are not only compatible but actually
quantitatively supported by coupling unification and GUT's. Another great asset of SUSY GUT's
is that proton decay is much slowed down with respect to the non SUSY case. While SUSY and GUT's go very well together, in all other frameworks that we have mentioned there is a nearby cut-off where the theory becomes non perturbative and there is no transparency up to the GUT scale. Also, the problem of adequately suppressing neutrino masses from contributions at the cut-off scale M through effective operators of the form $O_5\sim (\nu H)^T \frac{\lambda}{M}(\nu H)$ is present in all the alternative scenarios. SUSY has an excellent dark matter candidate: the neutralino. In fact, we know by now that most of the (flat) Universe is not made up of atoms [M. Turner]: $\Omega_{tot} \sim 1$, $\Omega_{baryonic} \sim 0.044$, $\Omega_{matter} \sim 0.27$, where $\Omega$ is the ratio of the density to the critical density. Most is Dark Matter (DM) and Dark Energy (DE). We also know that most of DM must be cold (non relativistic at freeze-out) and that significant fractions of hot DM are excluded. Neutrinos are hot DM (because they are ultrarelativistic at freeze-out) and indeed are not much cosmo-relevant: $\Omega_{\nu} \lappeq 0.015$. Identification of DM is a task of enormous importance for both particle physics and cosmology. If really neutralinos are the main component of DM they will be discovered at the LHC and this will be a great service of particle physics to cosmology.

It is really remarkable that all we know about neutrino masses is well in harmony with the idea and the mass scale of GUT's. As a consequence neutrino
masses have added phenomenological support to this beautiful idea and to the models of physics beyond the Standard Model
that are compatible with it. In particular, if we consider the main classes of new physics that are currently
contemplated, like supersymmetry, technicolour, large extra dimensions at the TeV scale, little Higgs models etc, it is
clear that the first one is the most directly related to GUT's. SUSY offers a well defined model computable up to the GUT
scale and is actually supported by the quantitative success of coupling unification in SUSY GUT's. For the other examples
quoted all contact with GUT's is lost or at least is much more remote. In this sense neutrino masses fit particularly well
in the SUSY picture that sofar remains the standard way beyond the Standard Model. 

\section{Baryogenesis through leptogenesis}

Another big plus that came out  of neutrino physics is the elegant picture of baryogenesis (BG) through leptogenesis (LG) (after LEP has disfavoured BG at the weak scale). In the Universe we observe an apparent excess of baryons over antibaryons. It is appealing that one can explain the
observed baryon asymmetry by dynamical evolution (BG) starting from an initial state of the Universe with zero
baryon number.  For BG one needs the three famous Sakharov conditions: B violation, CP violation and no thermal
equilibrium. In the history of the Universe these necessary requirements can have occurred at different epochs. But the asymmetry generated by one epoch could be erased at following epochs if not protected by some dynamical
reason. In principle these conditions could be verified in the SM at the electroweak phase transition. B is violated by
instantons when kT is of the order of the weak scale (but B-L is conserved), CP is violated by the CKM phase and
sufficiently marked out-of- equilibrium conditions could be realized during the electroweak phase transition. So the
conditions for baryogenesis  at the weak scale superficially appear to be present in the SM. However, a more quantitative
analysis
shows that baryogenesis is not possible in the SM because there is not enough CP violation and the phase
transition is not sufficiently strong first order, unless
$m_H<80~{\rm GeV}$, which is by now completely excluded by LEP. In SUSY extensions of the SM, in particular in the MSSM,
there are additional sources of CP violation and the bound on $m_H$ is modified by a sufficient amount by the presence of
scalars with large couplings to the Higgs sector, typically the s-top. What is required is that
$m_h\sim 80-110~{\rm GeV}$, a s-top not heavier than the top quark and, preferentially, a small
$\tan{\beta}$. But also this possibility has by now become at best marginal with the results from LEP2.
If BG at the weak scale is excluded by the data it can occur at or just below the GUT scale, after inflation.
But only that part with
$|{\rm B}-{\rm L} |>0$ would survive and not be erased at the weak scale by instanton effects. Thus BG at
$kT\sim 10^{10}-10^{15}~{\rm GeV}$ needs B-L violation at some stage like for $m_\nu$ if neutrinos are Majorana particles.
The two effects could be related if BG arises from LG then converted into BG by instantons. Recent results on neutrino masses are compatible with this elegant possibility [W. Buchmuller]. In particular the bound was derived on light neutrino masses $m_i \lappeq 0.1$ eV. This bound can be somewhat relaxed only for degenerate neutrinos but still remains compatible with the results from $\nu$ oscillations. Thus the case
of BG through LG has been boosted by the recent results on neutrinos.

\section{The Vacuum Energy Mystery}

Finally, we stress the importance of  the cosmological constant or vacuum energy problem [M. Turner]. The exciting recent results
on cosmological parameters, culminating with the precise WMAP measurements, have shown that vacuum energy accounts
for about 2/3 of the critical density: $\Omega_{\Lambda}\sim 0.65$, Translated into familiar units this means for the energy
density $\rho_{\Lambda}\sim (2~10^{-3}~eV)^4$ or $(0.1~mm)^{-4}$. It is really interesting (and not at all understood)
that $\rho_{\Lambda}^{1/4}\sim \Lambda_{EW}^2/M_{Pl}$ (close to the range of neutrino masses). It is well known that in
field theory we expect $\rho_{\Lambda}\sim \Lambda_{cutoff}^4$. If the cut off is set at $M_{Pl}$ or even at $0(1~TeV)$
there would an enormous mismatch. In exact SUSY $\rho_{\Lambda}=0$, but SUSY is broken and in presence of breaking 
$\rho_{\Lambda}^{1/4}$ is in general not smaller than the typical SUSY multiplet splitting. Another closely related
problem is "why now?": the time evolution of the matter or radiation density is quite rapid, while the density for a
cosmological constant term would be flat. If so, them how comes that precisely now the two density sources are
comparable? This suggests that the vacuum energy is not a cosmological constant term, buth rather the vacuum expectation
value of some field (quintessence) and that the "why now?" problem is solved by some dynamical mechanism. 

Clearly the cosmological constant problem poses a big question mark on the relevance of naturalness as a criterion also for the hierarchy problem: how we can trust that we need new physics close to the weak scale out of naturalness if we have no idea for the solution of the much more serious naturalness problem posed by the cosmological constant? The common answer is that the hierarchy problem is formulated within a well defined field theory context while the cosmological constant problem makes only sense within a theory of quantum gravity, that there could be modification of gravity at the sub-eV scale, that the vacuum energy could flow in extra dimensions or in different Universes and so on. At the other extreme is the possibility that naturalness is misleading. Weinberg has pointed out \cite{two} that the observed order of magnitude of $\Lambda$ can be successfully reproduced as the one necessary to allow galaxy formation in the Universe. In a scenario where new Universes are continuously produced we might be living in a very special one (largely fine-tuned) but the only one to allow the development of an observer. One might then argue that the same could in principle be true also for the Higgs sector. Recently it was suggested \cite{three} to abandon the no-fine-tuning assumption for the electro-weak theory, but require correct coupling unification, presence of dark matter with weak couplings and a single scale of evolution from the EW to the GUT scale. A "split SUSY" model arises as a solution with a fine-tuned light Higgs and all SUSY particles heavy except for gauginos, higgsinos and neutralinos, protected by chiral symmetry. In the same spirit also a two scale non-SUSY GUT (like SO(10, for example) could work, with axions as cold dark matter. In conclusion, it is clear that naturalness is normally a good heuristic principle in physics but one cannot prove its necessity in general.

\section{The $\nu$-Mixing Matrix}

If we take maximal $s_{23}$ and keep only linear terms in $u=  s_{13}e^{i\varphi}$, from experiment we find the following
structure of the
$U_{fi}$ ($f=e$,$\mu$,$\tau$, $i=1,2,3$) mixing matrix, apart from sign convention redefinitions: 
\bea  
&&U_{fi}= \nonumber\\
&&\left(\matrix{ c_{12}&s_{12}&u \cr  -(s_{12}+c_{12}u^*)/\sqrt{2}&(c_{12}-s_{12}u^*)/\sqrt{2}&1/\sqrt{2}\cr
(s_{12}-c_{12}u^*)/\sqrt{2}&-(c_{12}+s_{12}u^*)/\sqrt{2}&1/\sqrt{2}     } 
\right), \nonumber\\
\label{ufi1}
\eea
 Experimentally for the absolute values of the entries we approximately have:
\beq  U_{fi}= 
\left(\matrix{0.84&0.54&<0.25 \cr 0.44&0.56&0.71\cr
0.32&0.63&0.71     } 
\right) ~~~~~,
\label{ufiexp}
\eeq 
Note that $\theta_{12}$ is close to $\pi/6$ (for  $s_{12}=1/\sqrt{3}$ and $u=0$ we have the so-called tri-bimaximal mixing pattern, with the entries in the second column all equal to $1/\sqrt{3}$ in absolute value). Given the observed frequencies and  the notation $\Delta m^2_{sun}\equiv \vert\Delta m^2_{12}\vert$,
$\Delta m^2_{atm}\equiv \vert\Delta m^2_{23}\vert$ with
$\Delta m^2_{12}=\vert m_2\vert^2-\vert m_1\vert^2 > 0$ and $\Delta m^2_{23}= m_3^2-\vert m_2\vert ^2$, there
are three possible patterns of mass eigenvalues:
\bea 
&&{\tt{Degenerate}}:  |m_1|\sim |m_2| \sim |m_3|\gg |m_i-m_j|\nonumber\\ 
&&{\tt{Inverted~hierarchy}}:  |m_1|\sim
|m_2| \gg |m_3| \nonumber\\ 
&&{\tt{Normal~hierarchy} }:  |m_3| \gg |m_{2,1}|\nonumber\\
\label{abc}
\eea  
Models based on all these patterns have been proposed and studied and all are in fact viable at present [F. Feruglio].

The detection of neutrino-less double beta decay would offer a way to possibly disintangle the 3 cases.  The quantity which is bound by experiments
is the 11 entry of the
$\nu$ mass matrix, which in general, from $m_{\nu}=U^* m_{diag} U^\dagger$, is given by :
\bea 
\vert m_{ee}\vert~=\vert(1-s^2_{13})~(m_1 c^2_{12}~+~m_2 s^2_{12})+m_3 e^{2 i\phi} s^2_{13}\vert,
\nonumber\\
\label{3nu1gen}
\eea
Starting from this general formula it is simple to
derive the following bounds for degenerate, inverse hierarchy or normal hierarchy mass patterns,.
\begin{itemize}
\item[a)]  Degenerate case. If $|m|$ is the common mass and we take $s_{13}=0$, which is a safe
approximation in this case, because $|m_3|$ cannot compensate for the smallness of $s_{13}$, we have
$m_{ee}\sim |m|(c_{12}^2\pm s_{12}^2)$.  Here the phase ambiguity has been reduced to a sign ambiguity which is sufficient
for deriving bounds.  So, depending on the sign we have
$m_{ee}=|m|$ or
$m_{ee}=|m|cos2\theta_{12}$. We conclude that in  this case $m_{ee}$ could be as large as the present experimental limit
but should be at least of
order $O(\sqrt{\Delta m^2_{atm}})~\sim~O(10^{-2}~ {\rm eV})$ unless the solar angle is practically maximal, in which case
the minus sign option can be arbitrarily small. But the experimental 2-$\sigma$ range of the solar angle does not
favour a cancellation by more than a factor of 3.
\item[b)]  Inverse hierarchy case. In this case the same approximate formula $m_{ee}=|m|(c_{12}^2\pm s_{12}^2)$ holds 
because $m_3$ is small and $s_{13}$ can be neglected. The difference is that here we know that $|m|\approx 
\sqrt{\Delta m^2_{atm}}$ so that $\vert m_{ee}\vert<\sqrt{\Delta m^2_{atm}}~\sim~0.05$ eV. At the same time,
since a full cancellation between the two contributions cannot take place, we expect 
$\vert m_{ee}\vert > 0.01$ eV.

\item[c)]  Normal hierarchy case. Here we cannot in general neglect the $m_3$ term. However in this case $\vert
m_{ee}\vert~\sim~
\vert\sqrt{\Delta m^2_{sun}}~ s_{12}^2~\pm~\sqrt{\Delta m^2_{atm}}~ s_{13}^2\vert$ and we have the bound 
$\vert m_{ee}\vert <$ a few $10^{-3}$ eV.
\end{itemize}
Recently evidence for $0\nu \beta \beta$ was claimed [H.V. Klapdor-Kleingrothaus] at the 4.2-$\sigma$ level corresponding to
$\vert m_{ee}\vert\sim (0.2\div 0.6)~ {\rm eV}$ ($(0.1\div 0.9)~ {\rm eV}$ in a more conservative estimate of the involved nuclear matrix elements). If confirmed this would rule out cases b) and c) and point to case a) or to models
with more than 3 neutrinos.

After KamLAND, SNO and WMAP not too much hierarchy is measured for neutrino masses: $r = \Delta m_{sol}^2/\Delta m_{atm}^2 \sim 1/35$. Precisely at $3\sigma$: $0.018 \lappeq r \lappeq 0.053$. Thus, for a hierarchical spectrum, $m_2/m_3 \sim \sqrt{r} \sim 0.2$, which is comparable to the Cabibbo angle $\lambda_C \sim 0.22$ or $\sqrt{m_{\mu}/m_{\tau}} \sim 0.24$. This suggests that the same hierarchy parameters are at work for quark, charged lepton and neutrino mass matrices. This in turn suggests that, in absence of some special dynamic reason, we do not expect a quantity like $\theta_{13}$ to be too small. Note that sofar something like $\sin{\theta_{13}} \sim 1/2\sin{\theta_{12}}$ is not excluded at present. 

Pending the solution of the existing experimental ambiguities a variety of theoretical models of neutrino masses and
mixings are still conceivable. Among 3-neutrino models the normal hierarchy option appears to us as the most
straightforward and flexible framework. In particular the large atmospheric mixing can arise from lopsided matrices. Then the
observed frequencies and the large solar angle can also be obtained without fine tuning in models where the 23 subdeterminant
is automatically suppressed. 

The fact that some neutrino mixing angles are large and even
nearly maximal, while surprising at the start, was eventually found to be well compatible with a unified picture of quark
and lepton masses within GUTs. The symmetry group at
$M_{GUT}$ could be either (SUSY) SU(5) or SO(10)  or a larger group. For example, models based on anarchy, semianarchy, inverted hierarchy or normal hierarchy can all be naturally
implemented  by simple assignments of U(1)$_{\rm F}$ horizontal charges in a semiquantitative unified
description of all quark and lepton masses in SUSY SU(5)$\times$ U(1)$_{\rm F}$. Actually, in this context, if one adopts
a statistical criterium, hierarchical models appear to be preferred over anarchy and among them normal hierarchy appears the most likely [F. Feruglio]. Note that in almost all of the existing models of neutrino mixings the 
atmospheric angle is large but not  maximal. If it experimentally 
turned out that indeed $\theta_{23}$ is maximal with good accuracy then 
very special classes of models would be selected.

\section{Future Perspective}

\begin{figure}[htb]
\begin{center}
\includegraphics[width=8cm]{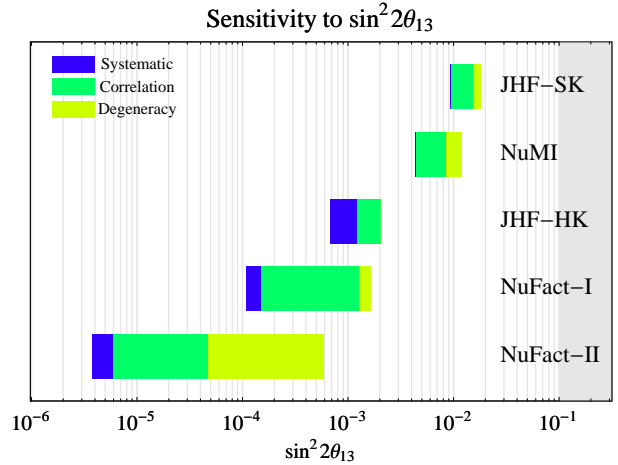}
\caption{Expected sensitivity to $\sin^2\theta_{13}$ of future experiments (courtesy of M. Lindner)}
\end{center}
\end{figure}

The goals of future experiments on neutrino masses and mixings that have been discussed in detail at this Conference are:
\begin{itemize}
\item[a)] Confirm or reject LSND. This is in progress by MiniBoone [S. Brice] which is taking data and will release its first results in 2005. \item[b)] Measure $\theta_{13}$ (in preparation: MINOS [M. Thompson], reactors [L. Oberauer], JPARC [Y. Hayato]....). In Fig.~4 the potential reach of different future experiments is shown (courtesy of M. Lindner).

 \item[c)] Detect $\nu_{\tau}$ in $\nu_{\mu} \leftrightarrow \nu_{\tau}$ (in preparation: OPERA [D. Autiero], ICARUS [A. Bueno]) \item[d)] How close to maximal is $\theta_{23}$? \item[e)] Determine the sign of $\Delta m_{23}^2$ to distinguish between normal or inverse hierarchy. \item[f)] Go after CP violation in the neutrino sector ( the last two items are for long baseline experiments  and $\nu$ factories [A. Blondel, A. Tonazzo, T. Kobayashi, M. Mezzetto]). \item[g)] Improve the sensitivity to $0\nu\beta\beta$ (CUORE [E. Fiorini], GENIUS  [H.V. Klapdor-Kleingrothaus] ,.... [F. Avignone]). 
\end{itemize}

To this list we can add related matters like \begin{itemize} \item[h)] Studies of cosmic neutrinos  [K. Mannheim] (Amanda [K. Woschnagg], ANTARES [G. Anton], Auger [E. Zas], Baikal [Zh. Dhzilkibaev], EUSO [S. Bottai], Icecube [O. Botner], NEMO [P. Piattelli], Nestor [S.E. Tsamarias]....). \item[i)] Lepton flavour violating processes ($\mu \rightarrow e\gamma$,....) [M. Aoki, C. Savoy], lepton moments [H. Wong] etc. \item[l)] Proton decay [C. K. Jung, L. Sulak].\end{itemize} 

It is clear that there is plenty of work for decades. I find that, beyond the immediate future, while Japan has a well defined roadmap (JPARC is on its way, funding has been allocated and there is planned neutrino physics for 2009), in Europe and the US there are many ambitious ideas, schemes, sites that are proposed and discussed, in particular at this Conference, but no convergence is in sight and, most important, no funding sofar. I really hope that this situation will soon improve. However, it must be kept in mind that all projects for 2010 and after must be compared with the results that, by then, will presumably start coming out from Japan.

Last, not least, as the last speaker, in behalf of all the participants, I would like to thank the Organizers for this perfect Conference. The College de France  is a great, comfortable, ideally located facility and Paris is one of the most attractive cities in the world, so that all of us have spent here a beautiful and most productive week.

\vfill
\end{document}